\documentclass[runningheads]{llncs}
\usepackage{graphicx}
\usepackage{array}
\newcolumntype{C}[1]{>{\centering\arraybackslash}m{#1}}
\usepackage{amsfonts}
\usepackage{amsmath}
\begin{document}
\title{Spatial-And-Context aware (\textbf{SpACe}) ``virtual biopsy'' radiogenomic maps to target tumor mutational status on structural MRI}

\author{Marwa Ismail\inst{1},
Ramon Correa\inst{1}, 
Kaustav Bera\inst{1},
Ruchika Verma\inst{1},
Anas Bamashmos\inst{2},
Niha Beig\inst{1},
Jacob Antunes\inst{1},
Prateek Prasanna\inst{1},
Volodymyr Statsevych\inst{1},
Manmeet Ahluwalia\inst{3},
Pallavi Tiwari\inst{1}}
\authorrunning{F. Author et al.}
%
\institute{Case Western Reserve University, Cleveland, OH, USA \and
Imaging Institute, Cleveland Clinic, Cleveland, OH, USA \and
Brain Tumor and Neuro-Oncology Center, Cleveland Clinic, Cleveland, OH, USA}

\maketitle              
\begin{abstract}
With growing emphasis on personalized cancer-therapies, radiogenomics has shown promise in identifying target tumor mutational status on routine imaging (i.e. MRI) scans. These approaches largely fall into two categories: (1) deep-learning/radiomics (context-based) that employ image features from the entire tumor to identify the gene mutation status, or (2) atlas (spatial)-based to obtain likelihood of gene mutation status based on population statistics. While many genes (i.e. EGFR, MGMT) are spatially variant, a significant challenge in reliable assessment of gene mutation status on imaging is the lack of available co-localized ground truth for training the models. 
We present Spatial-And-Context aware (SpACe) "virtual biopsy" maps that incorporate context-features from co-localized biopsy site along with spatial-priors from population atlases, within a Least Absolute Shrinkage and Selection Operator (LASSO) regression model, to obtain a per-voxel probability of the presence of a mutation status ($M^{+}$ vs $M^{-}$). We then use probabilistic pair-wise Markov model to improve the voxel-wise prediction probability. We evaluate the efficacy of SpACe maps on MRI scans with co-localized ground truth obtained from biopsy, to predict the mutation status of 2 driver genes in Glioblastoma (GBM): (1) $EGFR^{+}$ versus $EGFR^{-}$, (n=91), and (2) $MGMT^{+}$ versus $MGMT^{-}$, (n=81). When compared against state-of-the-art deep-learning (DL) and radiomic models, SpACe maps obtained training and testing accuracies of 90\% (n=71) and 90.48\% (n=21) in identifying EGFR amplification status, compared to 80\% and 71.4\% via radiomics, and 74.28\% and 65.5\% via DL. For MGMT methylation status, training and testing accuracies using SpACe were 88.3\% (n=61) and 71.5\% (n=20), compared to 52.4\% and 66.7\% using radiomics, and 79.3\% and 68.4\% using DL. Following validation, SpACe maps could provide surgical navigation to improve localization of sampling sites for targeting of specific driver genes in cancer.

\keywords{Tumor mutation status \and Radiogenomics \and spatial prior}
\end{abstract}
\section{Introduction}
 With treatments for solid cancers transitioning towards personalized therapy, mutational profiling for identifying target gene mutation status or gene amplification status using tissue biopsies is becoming \textit{status-quo} for most cancers. However, a significant challenge in reliable assessment of gene mutation or amplification status, is the underlying genomic heterogeneity which makes it challenging to identify the ``true'' mutational status based on random tissue sampling~\cite{koljenovic2002discriminating}. 
 Multiple studies have shown that certain gene mutations (e.g. MGMT promoter methylation, EGFR) have varying expressions across different parts of the tumor or between primary or secondary metastatic sites~\cite{qazi2017intratumoral,della2012mgmt}. There is hence a need for developing ``virtual biopsy'' techniques on imaging that can comprehensively capture the gene mutation heterogeneity of solid tumors, and potentially assist in surgical navigation to identify sampling sites for biopsy targeting.

The field of radiogenomics has provided a surrogate mechanism to predict gene mutational status on routine imaging (i.e. MRI) by training machine-learning models. Most of the existing radiogenomic models fall in two categories:  (1) deep learning~\cite{chang2018deep,korfiatis2017residual,li2017deep}/radiomics~\cite{li2018multiregional,parker2016intratumoral,french2019defining}, and (2) atlas-based probabilistic approaches~\cite{ellingson2013probabilistic,bilello2016population}. 
In the absence of co-localized biopsy sites on MRI,  deep learning/radiomic approaches employ features from the entire tumor to predict the gene mutational status.  In contrast, atlases-based approaches obtain the likelihood of the mutational status of driver genes such as MGMT and EGFR at different spatial locations by creating probabilistic radiographic atlases obtained from a large population. 
These population-based approaches however do not leverage any tumor-specific information in the model.

In this work, we present the first-attempt at creating ``virtual biopsy'' radiogenomic maps for predicting gene mutational status on MRI, by combining two complementary attributes that capture mutational heterogeneity at: (1) population-level via \textit{spatial-priors} for presence or absence of mutation status ($M^{+}, M^{-}$) using probabilistic atlases from a retrospective cohort, and (2) local tumor-level by incorporating \textit{context-priors} that capture mutational heterogeneity via radiomic attributes obtained from a stereotactically co-localized biopsy site within the tumor. The  spatial and context priors are combined within a Least Absolute Shrinkage and Selection Operator (LASSO) regression model to obtain a per-voxel probability of the likelihood of increased expression of the gene mutation ($M^{+}, M^{-}$) at that location. The prediction probabilities obtained for every voxel are further improved using probabilistic pairwise Markov models. In this work, we evaluate these Spatial and Context Aware (SpACe) maps in the context of two problems in Glioblastoma (GBM): (1) predicting EGFR status (amplified ($EGFR^{+}$), non-amplified ($EGFR^{-}$)), and (2) predicting MGMT status (methylated ($MGMT^{+}$), non-methylated ($MGMT^{-}$)), from routine MRI scans. The pipeline of the entire workflow is illustrated in Figure 1.

\section{Methods}
\subsection{Notation}
We deﬁne an image scene $I$ as $I = (C,f)$, where $I$ is a spatial grid $C$ of voxels $c \in C$, in a 3D space, $\mathbb{R}^3$. Each voxel, $c \in C$ is associated with an intensity value $f(c)$. $I_B$ represents the co-localized biopsy location on MRI scans, such that $I_B \subset I$. 
 $\mathbb{F}(c)$ denotes the feature set obtained for every $c \in C_B$.  For gene $M$, $M^{+}$ defines mutated/methylated, while $M^{-}$ defines non-mutated/unmethylated.

\begin{figure}[tb]
\includegraphics[height=11.7cm, width = 11.7cm]{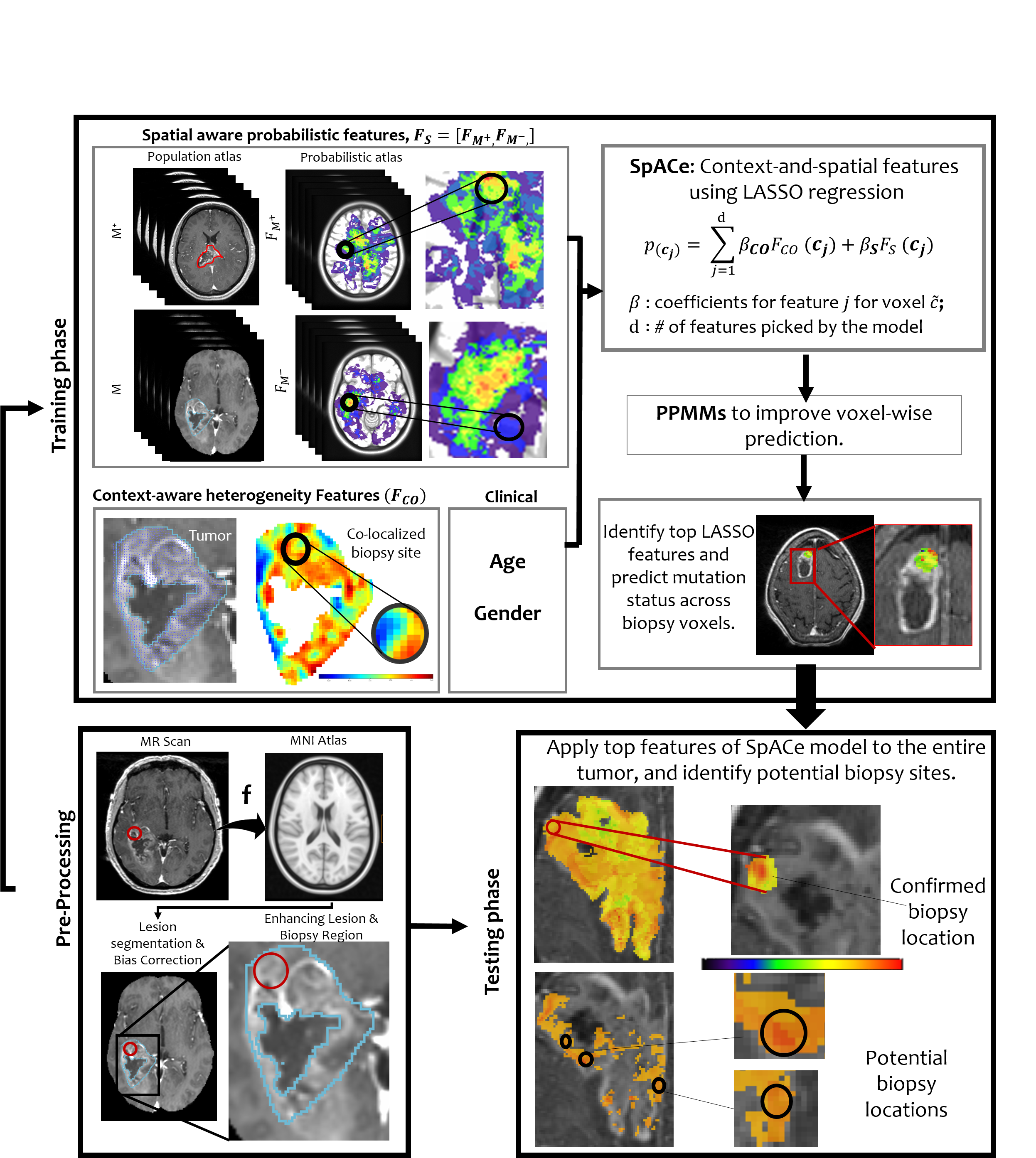}
\caption{Fig. 1: Overview of the workflow of \textbf{SpACe} to create ``virtual biopsy'' maps.} \label{fig1}
\end{figure}

\subsection{Computing context-aware mutational heterogeneity from stereotactic biopsy locations ($\mathbb{F_{CO}}$)}
\label{sec:context}
We define ``context'' as local heterogeneity attributes computed from the co-localized biopsy site on imaging, using radiomic features including Haralick features (capture image heterogeneity~\cite{haralick1973textural}), Gabor features (capture structural details at different orientations and scales), Laws (capture spots and ripples-like patterns), and CoLlAGe features (capture localized gradient orientation changes~\cite{prasanna2016co}). Specifically, for every $c$ $\in$  $C_B$, we extract a set of 3D radiomic features (i.e. Haralick, Gabor, Laws, CoLlAGe). We define $\mathbb{F}_{\theta}^k (c)$, where $\theta$ is the type of feature family (e.g. Haralick, Gabor),  and $k \in \{1,...,n\}$, where $n$ is the number of feature attributes for every feature family. 
Feature pruning is then conducted on the extracted features using Spearman’s correlation metric, to eliminate redundant features. The pruned ``context-aware'' features (152 for EGFR cohort, 149 for MGMT cohort) are finally aggregated into one feature descriptor $\mathbb{F_{CO}}$.

\subsection{Computing spatially-aware priors ($\mathbb{F_{S}}$) for likelihood of gene mutation status ($M^{+}, M^{-}$) using probabilistic atlases}
\label{sec:spatial}
Using the lesion segmentation obtained for every patient in the training set, two different population atlases for gene $M$ are constructed using subjects that belong to either $M^{+}$ or $M^{-}$. This is done to quantify the frequency of occurrence of every voxel across $M^{+}$ and $M^{-}$, and compute voxel-wise probability values, $P_{w}(c)$, $w \in$ ($M^{+}, M^{-}$). All scans need to first be registered to an isotropic reference atlas (i.e. MNI152; Montreal Neurological Institute).  The intensity values are then averaged across $c \in C$ across all the annotated binary images of all patients involved in the study. This means that for $c \in C$, two probability values  from these two atlases could be obtained, that characterize the probability of a voxel $c$ being $M^{+}$ or $M^{-}$. The 2 probability values ($P_{M^+}, P_{M^-}$) for every voxel $c \in C$ are finally aggregated in the spatial feature descriptor $\mathbb{F}_{S}=[P_{M^+}, P_{M^-}]$. 

\subsection{Creating SpACe maps for predicting voxel-wise mutational heterogeneity in the tumor}
\label{sec:space}
In order to obtain a voxel-wise prediction $p(c)$ of the gene mutation status, the context descriptor ($\mathbb{F_{CO}}$), spatial descriptor ($\mathbb{F_{S}}$), age ($\mathbb{F_{A}}$), and gender ($\mathbb{F_{G}}$) of every patient in the training set, are incorporated within a LASSO model~\cite{tibshirani1996regression}. LASSO model is selected to obtain the probability score using a parsimonious feature set by  utilizing its capability in reducing variance when shrinking features, while simulteneously not increasing the bias. We designed the LASSO model to perform regularization of feature parameters as follows:
[$\hat{\beta}$] = $argmin \{|y - F_{CO}\beta_{CO}|^2 + \lambda_{CO}$ $|\beta_{CO}| + |y - F_{S}\beta_{S}|^2 + \lambda_{S} |\beta_{S}| + |y - F_{A}\beta_{A}|^2 + \lambda_{A} |\beta_{A}|+ |y - F_{G}\beta_{G}|^2 + \lambda_{G} |\beta_{G}|\}$, where $[\hat{\beta}] = \{\hat{\beta_1},...,\hat{\beta_d}\}$ is the shrinked set of $d$ coefficients obtained after regularization, $y\in [M^{+}, M^{-}]$, and $\lambda$ is the penalty term. 
The voxel-wise probability is then computed as the weighted sum of the selected features for the set of coefficients $[\hat{\beta}]$, as follows:
   $p_{SpACe}(c)  = \sum_{j=1}^{d} \hat{\beta}_{j}\mathbb{F}_{j}(c)$,
 where $j \in \{1,...,d\}$, and $d$ is the number of features selected by LASSO. After obtaining the probabilistic map for every $c \in C_B$ , we incorporate probabilistic pairwise Markov models (PPMMs) to improve voxel-wise gene mutation prediction~\cite{monaco2010high}. PPMMs are adopted from Markov Random Fields, through formulating priors in terms of probability density functions, hence allowing the creation of more robust prediction models. The input to this model is the voxel-wise probability values obtained from LASSO model. Interaction between neighboring sites is then modelled, to improve voxel-wise probability scores, and to finally obtain $\mathbb{F}_{SpACe}$ maps. 
\subsection{Applying SpACe maps on testing sets for predicting voxel-wise mutational heterogeneity within the tumor}

The top features selected on the training set are applied to the entire tumor on the test set, for obtaining voxel-wise probabilities for predicting the mutation status. For the purpose of computing accuracy of our model, we predict the mutation status [$M^{+}$, $M^{-}$] based on pooled probability values for an already known biopsy site, and compare the prediction with the known mutation status. As an additional qualitative analysis, we threshold the probability values obtained from the entire tumor (threshold obtained empirically), followed by connected component analysis and PPMM, to obtain 2-3 hot-spots of high probability mutation sites. These hot-spots prospectively could be used to drive surgical navigation as potential sites for biopsy localization.

\section{Experimental Design}
\subsection{Data description and preprocessing}
We employed a unique retrospective  dataset of a total of 100 GBM patients who underwent CT-guided biopsy for disease confirmation, since surgical resection was not feasible (due to location or other clinical reasons) for these patients. Segmentation of the enhancing lesion was conducted by an experienced radiologist on the MR scans. The biopsy site was co-localized by co-registering CT images with the MRI scans, followed by expert evaluation for confirmation. All scans were then registered to an MNI152 atlas and then bias-corrected using N4 bias correction ~\cite{tustison2010n4itk}. These studies were then divided into two cohorts: (a) $S_1$: EGFR amplified ($EGFR^+$) versus non-amplified ($EGFR^-$) studies, and (2) $S_2$: MGMT methlated ($MGMT^+$) versus unmethylated ($MGMT^+$).  For $S_1$, we had a total of 91 subjects of which 70 were  used for training (35 amp, 35 non-amp), and the remaining 21 (6 $EGFR^+$, 15 $EGFR^-$) were used for validation. For $S_2$, of a total of 81 subjects, 60 (28 $MGMT^+$, 32 $MGMT^-$) were used for training , while 21 subjects were used for validation (5 $MGMT^+$, 16 $MGMT^-$). 

\subsection{Implementation details}
Two experiments were set-up using cohorts $S_1$ (\textbf{Experiment 1}: $EGFR^+$ versus $EGFR^-$) and $S_2$ (\textbf{Experiment 2}: $MGMT^+$ versus $MGMT^-$), respectively. For both experiments, we extracted a total of $316$ 3D context-features for every $c\in C_B$ (where $C_B$ was a 1-cm diameter sphere in our case), including 1 raw feature, 8 gray features, 13 gradient features, 26 Haralick features, 64 Gabor features, 152 Laws features, and 52 CoLlAGe features, extracted using 2 window sizes $w= 3 \times 3$ and $w=5\times 5$. These features are pruned as detailed in Section~\ref{sec:context} to obtain $\mathbb{F_{CO}}$. In addition, population atlases were constructed to quantify the frequency of occurrence of $EGFR^+$ versus $EGFR^-$ and $MGMT^+$ versus $MGMT^-$ as detailed in Section~\ref{sec:spatial}. Similarly, for both experiments, $\mathbb{F}_{SpACe}$ was created following 10 runs of 10-fold cross validation, as detailed in Section ~\ref{sec:space}. The median value of the probabilities across voxels of all subjects was used as a threshold to determine $M^+,M^-$ for every voxel. Finally, majority voting was used to obtain the mutation status for every biopsy site.    

\subsection{Comparative strategies}
In order to evaluate the efficacy of SpACe model, we compared our results with two state-of-the-art methods, radiomic-based, and deep-learning-based that employ features from the entire tumor to predict amplification/methylation status. For the radiomic- experiment ($\mathbb{F}_{Rad}$, a total of 316 radiomic features were extracted from the entire enhancing lesion of every subject (same attributes that were extracted from biopsy sites), and the feature vector was constructed from the 4 statistics: median, variance, skewness, and kurtosis values that were computed for every feature across all voxels for this patient, for total of 1264 features.  After feature pruning, 283 features were fed to the LASSO model to compute patient-wise scores that determined their gene status.  

For the DL approach to predict the mutation status, we used a deep residual neural network as described in~\cite{korfiatis2017residual}. ResNet has previously been used to predict EGFR and MGMT mutation in GBM and other cancers~\cite{korfiatis2017residual,xiong2019implementation}. 
Specifically, patches of size 128x128 were sampled from the center of the selected MRI slices and augmented using horizontal flips and random rotations to enlarge the limited training data. Following patch sampling, we trained separate deep Res-Net networks with 18 layers (ResNet-18) for the two experiments. In order to train the networks on MRI scans, we used pre-trained model on ImageNet and performed transfer learning using the sampled patches from MRI scans. We selected ResNet-18 because it removes the vanishing gradient problem and the network has several layers containing composite function of operations such as batch normalization (BN), convolution (Conv), rectified linear units (ReLU) and pooling for non-linear transformation of the input. We trained each model for 25 epochs with dropout of 0.2 to avoid overfitting. Models with minimum loss were locked down to test the patches obtained from the test set.

\begin{figure}[bt]
\includegraphics[width=\textwidth]{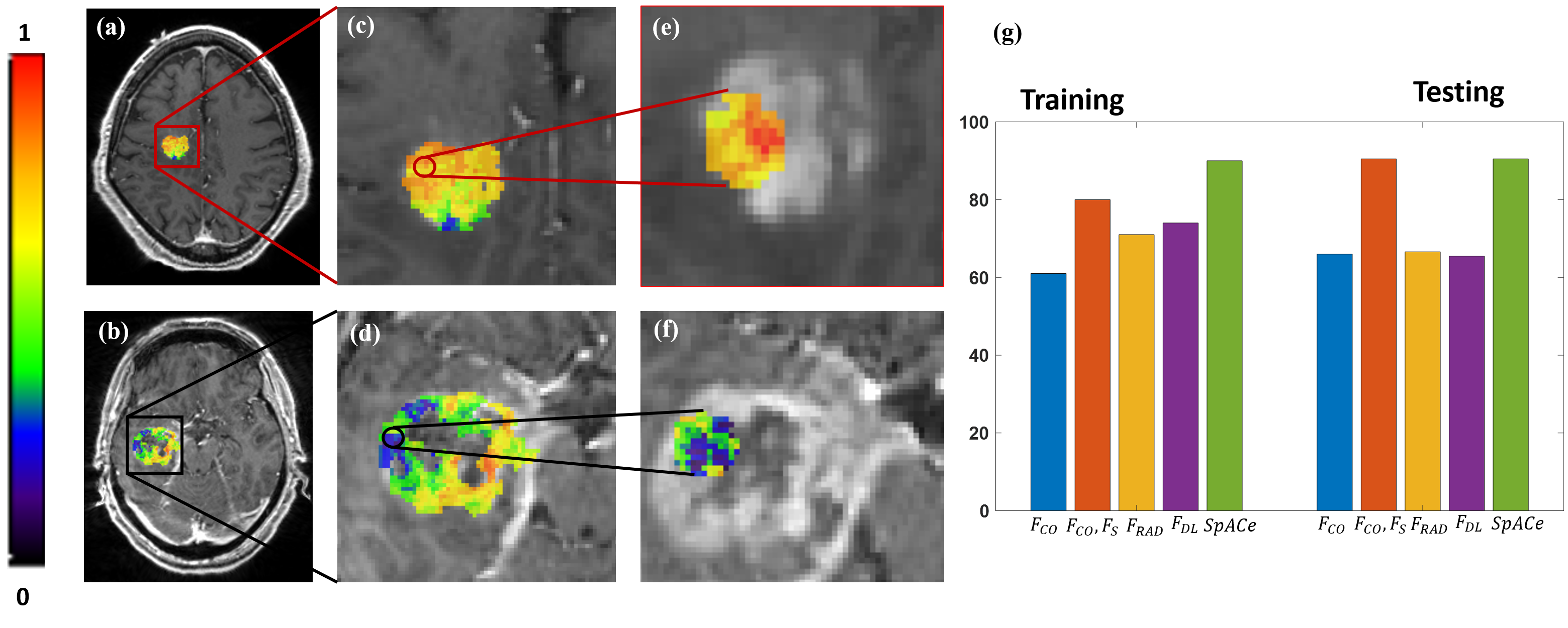}
\caption{$EGFR^+$ (a) and $EGFR^-$ (b) cases with voxel-wise probabilities calculated using SpACe maps. Heatmaps with voxel-wise probabilities for the entire tumor area for $EGFR^+$ (c) and $EGFR^-$ (d) are shown, where "red" represents amplified and "blue" represents un-amplified status. Confirmed biopsy sites are enclosed by a circle. (e), (f) show biopsy region heatmaps, which confirm the mutation status of the tumor. Tumor heatmaps in (c), (d) show other clusters that could be potential candidates for biopsy sites. The prediction accuracies for predicting mutation status in the two patients in (a) and (b) using SpACe were 92.5\% and 96\% respectively. (g) shows a bar graph with accuracies for both training and testing sets for $EGFR^+$ versus $EGFR^-$, using $\mathbb{F}_{CO}$, [$\mathbb{F}_{CO}$. $\mathbb{F}_{S}$], $\mathbb{F}_{Rad}$, $\mathbb{F}_{DL}$, and  $\mathbb{F}_{SpACe}$.} \label{fig2}
\end{figure}

\section{Results and Discussion}
\subsection{Experiment 1: Determining EGFR amplification status}
Using $\mathbb{F_{CO}}$ alone, training and testing accuracies were reported as 61.43\% and 66.67\% respectively. Combination of $\mathbb{F_{CO}}$ features with $\mathbb{F_{S}}$ features into LASSO model yielded training and testing accuracies of 80\% and 90.48\% respectively, using 8 $\mathbb{F_{CO}}$ (1 raw, 1 gray, 2 gradient, 1 Haralick, 3 Gabor) and 2 $\mathbb{F_{S}}$ features. This implies that incorporating $\mathbb{F_{S}}$ improved the model’s performance, rather than using $\mathbb{F_{CO}}$ alone. Next, we evaluated the efficacy of including $\mathbb{F_{CO}}$, $\mathbb{F_{S}}$, and clinical features ($\mathbb{F_{A}}$, $\mathbb{F_{G}}$) into our model to predict the mutation status. Clinical features did not improve accuracy of the model. PPMMs were then employed, and successfully corrected the amplification status for 7 subjects from the training set, yielding final training and testing accuracies of 90\% (1 amp, 6 non-amp subjects were misclassified out of 70) and 90.48\% (2 non-amp subjects are misclassified out of 21) respectively. Results on 2 different patients are illustrated in Figure 2.

Using radiomic features from the entire tumor to predict mutation status yielded training and testing accuracies of 80\% and 71.43\% respectively. Further, the Res-Net model to predict EGFR status yielded training and testing accuracies of 74.28\% and 65.52\%, significantly underperforming in comparison to the SpACe model. 

\begin{figure}[tb]
\includegraphics[width=\textwidth]{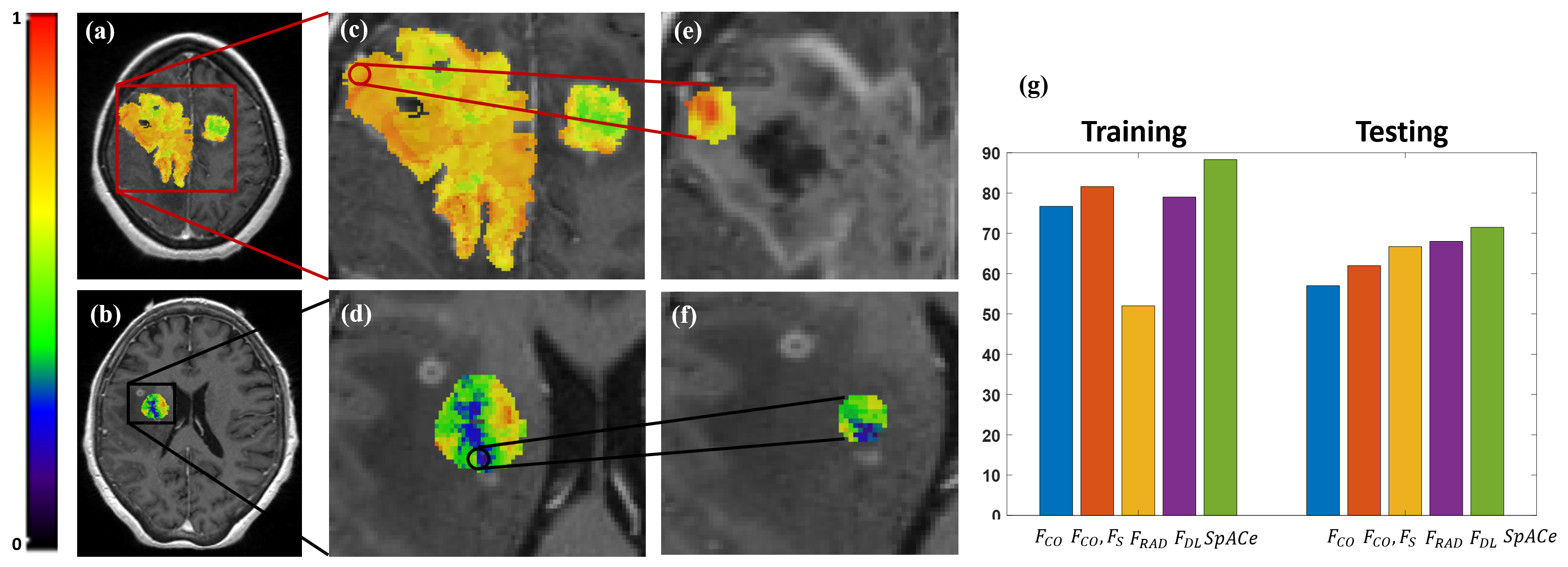}
\caption{$MGMT^+$ (a) and $MGMT^-$ (b) cases with voxel-wise probabilities calculated using SpACe maps. Heatmaps with voxel-wise probabilities for the entire tumor area for $MGMT^+$ (c) and $MGMT^-$ cases (d) are shown, where "red" represents methylated and "blue" represents unmethylated status. Confirmed biopsy sites are enclosed by a circle. (e), (f) show biopsy region heatmaps, which confirm the mutation status of the tumor. Tumor heatmaps in (c), (d) show other clusters that could be potential candidates for biopsy sites. The prediction accuracies for cases (a) and (b) using SpACe are 98\% and 99\% respectively. (g) shows a bar graph with accuracies for both training and testing sets using $\mathbb{F}_{CO}$, [$\mathbb{F}_{CO}$. $\mathbb{F}_{S}$], $\mathbb{F}_{Rad}$, $\mathbb{F}_{DL}$, and  $\mathbb{F}_{SpACe}$.} \label{fig3}
\end{figure}
\subsection{Experiment 2: Determining MGMT methylation status}
Using $\mathbb{F_{CO}}$ alone, the model achieved training and testing accuracies of 76.67\% and 57.14\% respectively. When combining $\mathbb{F_{CO}}$ features with $\mathbb{F_{S}}$ features into LASSO model, we obtained training and testing accuracies of 81.67\% and 61.9\% respectively, which implies that incorporating $\mathbb{F_{S}}$ improved the model’s performance, rather than using $\mathbb{F_{CO}}$ alone. Next, using $\mathbb{F_{CO}}$ , $\mathbb{F_{S}}$, and clinical features, the model picked a set of 12 features that included 8  $\mathbb{F_{CO}}$ features; 1 gray, 3 Haralick, and 4 Gabor features, in addition to $P_{MGMT^+}$, $P_{MGMT^-}$, $\mathbb{F_A}$, and $\mathbb{F_G}$. This model yielded training and testing accuracies of 83.3\% and 66.67\% respectively. Applying PPMMs for predicting methylation status on these results corrected the mutation status for 3 training subjects as well as 1 testing subject, with final accuracies of 88.3\% and 71.5\% respectively. Results on 2 different patients are illustrated in Figure 3.

When using radiomic features from the entire tumor to predict methylation status, training and testing accuracies were 76.67\% and 52.38\%. In addition, the DL model that was trained to predict methylation status gave training and testing accuracies of 79.37\% and 68.40\%, suggesting that results obtained using SpACe maps outperformed both comparative approaches.

\section{Concluding Remarks}
In this work, we presented the first-attempt at creating ``virtual biopsy'' radiogenomic maps for predicting gene mutational status on MRI, by combining two complementary attributes: (1) \textit{spatial-priors} for presence or absence of mutation status  via probabilistic atlases from a retrospective cohort, and (2) \textit{context-priors} to capture mutational heterogeneity using radiomic attributes obtained from a stereotactically co-localized biopsy site within the tumor. These spatial-and-context aware (SpACe) maps were evaluated in  the context of two experiments: predicting (1) EGFR amplification status, and (2) MGMT amplification status, on Glioblastoma. Our results demonstrated that SpACe outperformed state-of-the-art radiomic and deep learning approaches that were performed on  the entire tumor, instead of learning features from the co-localized biopsy site. The virtual biopsy maps created using SpACe could not only improve prediction of gene mutation status of the tumor, but could also serve as surgical navigation to guide potential biopsy sites for specific gene mutations.

%
%
%
%
%
 \bibliographystyle{splncs04}

 \bibliography{ref}

\end{document}